\documentstyle[preprint,aps,epsfig]{revtex} 
\begin{document}
\draft
\title{Tilt Grain-Boundary Effects in $S-$ and $D-$Wave Superconductors}
\author{J.J. Hogan-O'Neill$^{\dagger}$, A.M. Martin$^{\S}$ and 
James F. Annett$^{\dagger}$}
\address{$^\dagger$H.H. Wills Physics Laboratory, University of Bristol, 
Royal Fort, Tyndall Avenue, Bristol, BS8 1TL, U.K. }
\address{$^{\S}$D\'{e}partement de Physique Th\'{e}orique, 
Universit\'{e} de Gen\`{e}ve, 1211 Gen\`{e}ve 4, Switzerland}
\date{\today}
\maketitle
\begin{abstract}
We calculate the $s-$ and $d-$wave superconductor order parameter in 
the vicinity of a tilt grain boundary.  We do this self-consistently within
the Bogoliubov de Gennes equations, using a realistic microscopic model
of the grain boundary. We present the first self-consistent calculations
of supercurrent flows in such boundaries, obtaining the
current-phase characteristics of grain boundaries in both $s-$wave and $d-$wave 
superconductors. 
\end{abstract}

The debate over the superconducting order-parameter in the high $T_{c}$ 
superconductors (HTSC) has been strongly contested, but has now 
been settled in favour of a $d-$wave pairing state{\cite{annett1,annett2}}. 
The experiments by Tsuei {\it et al.}{\cite{Tsuei}} and Wollman {\it et al.}
{\cite{Wollman}} have been most conclusive, 
especially since they only depend on the 
phase of the order-parameter and not on the microscopic physics 
of the energy gap. 
Photoemission experiments{\cite{Ding}} and the temperature 
depedence of the penetration 
depth{\cite{Bonn}} also strongly support the $d-$wave picture. 
However, some controversy 
still exists concerning possible $s-d$ mixing in the cuprates. In 
particular an $s$-wave component has been demonstrated to be induced 
at interfaces{\cite{Sun,Kleiner,Covington}}.

In interpreting the Tsuei {\it et al.} experiments it is essential to
understand the superconducting characteristics of grain-boundary (GB) 
weak links in the cuprates. Understanding the effects of grain boundaries 
is also of importance for developing possible devices and  other applications 
of high $T_c$ superconductors. The 
Tsuei {\it et al.} experiments, especially the 
observations of $\pi$-junction behaviour, 
were consistent with the predictions of 
$d-$wave pairing interpreted using the Sigrist-Rice{\cite{Sigrist}} 
model for the dependence of the critical current, $I_ c$, 
on the grain boundary angles.  On the other hand the values of $I_c$ 
measured as a function of grain boundary angle show an almost 
exponential dependence on angle{\cite{Ivanov,Mannhart}, 
unlike the cosine predicted by the 
Sigrist-Rice formula for $d-$wave pairing. In 
trying to explain these findings a number of different models of the 
interfaces have been studied. Tanaka and collegues {\cite{Tanaka}} 
have looked at the $(100)$ and $(110)$ interfaces and also 
derived a Josephson current formula for $s/I/d$ and $d/I/d$ 
grain boundary structures. Barash {\it et al.}{\cite{Barash}} 
have considered the temperature dependence of the critical 
current in $d$-wave junctions. Zhang considered $0^o$, 
$45^o$, and $90^o$ junctions{\cite{Zhang}. Zhitomirsky 
and Walker {\cite{Zhit}} have also looked at the $(110)$ interface to study the 
quasiparticle spectra and zero energy states (ZES). Beltzig 
{\it et al.}{\cite{Belzig}} 
showed that an induced $s-$wave component existed on the (orthorhombic) $(110)$ boundary 
giving rise to a splitting in the ZES at a low enough temperature: the latter 
point they attribute to Time Reversal Symmetry Breaking (TRSB). The review 
on GBs by Prester{\cite{Prester}} also highlights the possibility of 
them behaving as though each were an individual Josephson Junction. 
Gurevich and Pashitskii argued that the near exponential 
depencence of $I_c$ on angle was due to the formation of an insulating 
layer at the grain boundary, assosciated with the dislocation 
cores{\cite{Gurevich}}.

In this letter we address the effects of grain boundaries in both
$s-$ and $d-$wave superconductors using a fully self-consistent solution
of the Bogoliubov de Gennes (BdG) equations. We adopt a
geometrically realistic model of the tilt grain boundary (GB), as shown in Fig.{\ref{fig1}}.  
By solving the BdG equations in real space
using the Recursion Method we are able to study such complex geometries,
unlike earlier calculations which were limited to either simpler interfaces
or planar junction models{\cite{Tang,Zhang,Otterlo}}.
In consequence we can determine how the superconducting order-parameter 
($\Delta$), the charge density ($n$), and the quasi-particle local density 
of states (LDOS, $n(E)$) are affected by our GB. Further, by solving the BdG 
equations in a self-consistent manner we can apply 
phase-differences in $\Delta$ across the boundary and calculate the resulting 
supercurrent. By calculating the maximum current across the boundary we 
determine the critical current of the system. 

For the purposes of this letter, we concentrate on the large angle 
grain-boundary ($53.1^o$), depicted in Fig.1. It consists of two square
lattices butted together at some angle of misorientation and linked via a 
{\it percolation site}. Being periodic in the $y$ direction 
we only have to undertake calculations for sites on two disimilar lines 
of atoms as shown in Fig. 1. For each line of sites we have    
to go 10 sites deep into the bulk before the order parameter
has recovered to its bulk value. 
For some bonds across the GB the interatomic spacing is less than the bulk 
spacing. For these bonds we assume that the value for the hopping between 
sites $i$ and $j$ ($t_{ij}$) can be 
calculated by assuming a simple linear form for the hopping-integral, 
i.e.
\begin{equation}
t_{ij}=-\frac{\sqrt{2}-r_{ij}}{\sqrt{2}-1}, 
\:\:\:\:\: 0\leq r_{ij}\leq \sqrt{2}
\end{equation}
and is zero otherwise. For the geometry of our GB (Fig.1) every 
site will have a {\it connectivity} of $4$. 
 
We consider the following two attractive Hubbard models:
\begin{equation}
H = \sum_{<i,j>,\sigma}t_{ij}c_{i\sigma}^\dagger c_{j\sigma}+ h.c. 
+U\sum_{i}n_{i\uparrow} n_{i\downarrow} \\
\end{equation}
\begin{equation}
H = \sum_{<i,j>,\sigma}t_{ij}c_{i\sigma}^\dagger c_{j\sigma}+ h.c. 
+U\sum_{<i,j>}n_{i} n_{j}
\end{equation}
with $U<0$ and $n_{i}=n_{i\uparrow}+n_{i\downarrow}$. Here $U$ is the 
usual BCS pairing-potential, defined as being a negative 
constant within a cutoff energy range of $\pm E_{c}$ either side 
of the Fermi energy, after which its value is zero. 
Eq.2 will be refered to as `local', giving rise to $s-$wave pairing 
and Eq.3 will be termed `non-local' giving $d-$wave pairing. 
By making the Bogoliubov canonical-
transformation we diagonalise the Hamiltonian and arrive at the 
Bogoliubov-de Gennes equations
\begin{equation}
\sum_{j} 
\left( \begin{array}{cc}
H_{ij} & \Delta_{ij}  \\
\Delta^{\star}_{ij} & 
-H^{\star}_{ij}
\end{array} \right)
\left( \begin{array}{c}
u_{j}^{n} \\
v_{j}^{n}
\end{array}
\right)
=
E_{n}
\left(
\begin{array}{c}
u_{i}^{n}\\
v_{i}^{n}
\end{array}
\right)
\end{equation}
where $H_{ii}=(- \mu + \frac{1}{2}U n_{ii})$ and $H_{ij}=t_{ij}$ (local), 
or, $H_{ii}=-\mu$ and $H_{ij}=t_{ij}+\frac{1}{2} U n_{ij}$  (non-local). 
Here $\mu$ is the chemical potential,  $u_i^n$ and $v_i^n$ are 
the particle and hole amplitudes on site $i$ associated with 
an eigenenergy $E_n$, and $n_{ij}$ is the appropriate charge density 
(defined below). To solve these equations we employ the Recursion 
Method{\cite{Haydock}}, and together with the 
methods employed in Martin and Annett {\cite{Martin}}, 
we obtain a matrix continued-fraction for the Green functions. 
This continued-fraction is evaluated exactly to 50 levels after which its 
elements vary in a predictable manner and therefore can be extrapolated for 
a futher 1500 levels say.

We are interested in evaluating the local quasi-particle density of 
states, the local and non-local charge densities 
($n_{ii}=\sum_{\sigma}\langle c_{i \sigma}^{\dagger}c_{i \sigma}\rangle$ 
and $n_{ij}=\sum_{\sigma}\langle c_{i \sigma}^{\dagger}c_{j \sigma}\rangle$ 
respectively), and the local and non-local order-parameters 
($\Delta_{ii}=U\langle c_{i \uparrow}c_{i \downarrow}\rangle$ 
and $\Delta_{ij}=U\langle c_{i \uparrow}c_{j \downarrow}\rangle$ respectively). 
These quantities may be found from 
the Green functions, expressions for which have already been given 
elsewhere {\cite{Martin}}. In the calculations that follow we make a BCS cutoff of 
$U=-3.5t$, have a temperature, $T$, of $T=0.01t$, and $E_{c}=3.0t$. 
Iterating the equations for charge-densities and order-parameters, with 
the BdG equations, we generate self-consistent solutions. So as not to 
direct the final SC solution into a local energy minimum, we set 
the order-parameter to zero at the beginning of the calculation on those sites 
closest in proximity to the boundary. We say 
self-consitency has been achieved when the Hartree-Fock term and the 
order-parameter change by less than a predefined margin between iterations. 
For the $s-$wave case, we can reach $0.5\%$ s.c. typically between 
10 or 20 iterations, whereas 
for the $d-$wave it usually takes over 200 iterations.

By imposing a phase-difference $\varphi$ in the 
order-parameter between the 
two bulk regions we may now generate current-flow across the GB. We 
initially make the Peierls substitution for 
$t_{ij}$, ($t_{ij}\rightarrow t_{ij}e^{-ieA_{ij}/\hbar}$, 
$A_{ij}$ being the integral of the vector-potential between sites 
$i$ and $j$), in Eq.2 and 3, and use the definition 
$I_{ij}= \langle \frac {\partial H}{\partial A_{ij}}\rangle$, 
to obtain {\cite{Miller}}
\begin{eqnarray}
I_{ij}&=&\frac {t_{ij}e}{\hbar} \Re \Big[ \frac {1} {\pi}
\int_{- \infty}^{+ \infty}\big[ G_{ij}(E+\imath \eta) \nonumber\\
&\;&\quad\quad-G_{ij}(E-\imath \eta)\big]e^{-ieA_{ij}/\hbar} f(E)dE
\Big]
\end{eqnarray}
$f(E)$ being the Fermi-Dirac function. By a suitable choice of gauge 
we can immediately set $A_{ij}=0$ everywhere.

When applying phase-differences in $\Delta$ across a GB it is instructive to 
note how $\Delta$ changes due to the self-consitency, and also observe how 
the local densities of states and the Hartree-Fock term alter. 
Fig.{\ref{fig2}} shows the results for the $s-$wave case. Figs.2(a) and 2(b) show the 
evolution of $|\Delta_{ii}|$ across the GB on both the 
lower (a) and upper (b) lines of SC sites at 
$\varphi=0^o$ (full line) and also $\varphi=180^o$ (dotted line). The GB 
obviously has a perturbing effect 
on the system but note that $|\Delta_{ii}|$ is not 
depressed on the percolation site ($x=0$) for $\varphi=0^o$. 
Infact $|\Delta_{ii}|$ is almost constant as $\varphi$ is varied; except for 
$\varphi=180^o$ where $|\Delta_{ii}|=0$ on the percolation site and is 
strongly reduced nearby. Note also the small Friedel-like oscillations in 
$|\Delta_{ii}|$ near the GB. Fig.2(c) shows 
how the order parameter phase, $arg(\Delta_{ii})$, varies as a 
function of $x$ co-ordinate through the GB for the $\varphi=30^o$ case. 
For the $180^o$ case we just 
have a step function with arbitrary phase associated with the percolation 
site since $|\Delta_{ii}|=0$. For all other phase 
differences, the phase as a function of $x$ goes as 
$(\varphi/2)\tanh(x/d)$ where $\varphi$ is the bulk phase difference 
across the GB and $d$ is a characteristic length scale which we find to 
be $d=3.5$ for our parameters. Also, this equation holds for both SC lines. 
Finally, in Fig.2(d) we show the local density of states on the 
percolation site at $\varphi=0^o$ (full line) and at $\varphi=180^o$ 
(dotted line). At $0^o$ we 
have the usual BCS-like gap around the Fermi energy. This persists through 
all our calculated phase-differences except at $180^o$ where the gap is 
suddenly filled in with various resonant states. Although not presented here 
we note that the Hartree-Fock term $\frac{1}{2}Un_{ii}$ shows 
the same (enhanced) values on both lines of SC sites at different 
phase-differences. 

Now consider the $d-$wave case. 
Because each site has four bonds with order parameters $\Delta_{ij}$ we can 
calculate the `net' $d-$wave contribution at a particular site by considering 
$\Delta_{i}^d=\sum_{j=1,4}(-1)^j \Delta_{ij}$
Similary, the extended$-s$ component 
is given by $\Delta_{i}^s=\sum_{j=1,4}\Delta_{ij}$ 
and will be finite near the boundary because of the broken crystallographic 
symmetry. A minor difficulty now arises when discussing the $d-$wave scenario. 
In the definition of $\Delta_{i}^d$ we must choose 
a convention for the direction of the positive and negative lobes of the 
$d-$wave function. For GB angles close to $45^o$, such as in Fig.1, there 
is an ambiguity in defining the relative orientations of the order parameter 
lobes either side of the GB. The convention we have chosen is indicated in 
Fig.1. In this case the current-phase relationship, $I(\varphi)$, is similar 
for both $s-$wave and $d-$wave cases. For this choice the $d-$wave order 
parameter is zero on the percolation site, $x=0$, for zero phase difference, 
Fig.{\ref{fig3}}(a), unlike the $s-$wave case of Fig.2(a). Using our definition 
of phase-differences we find a maximum $d-$wave contribution at $180^o$ 
on the lower line of SC sites (dotted line), which decreases with  
phase-difference, down to $0^o$ (full line) where $|\Delta^{d}_{x=0}|=0$. 
The upper line of SC sites (Fig.3(b)) shows the same 
qualitative form for $\Delta_{d}$ as for the 
local $s-$wave case. At $0^o$ we find maximum extended$-s$ component on the 
precolation site (full line in Fig.3(c)), decreasing with increasing 
phase-difference until at $180^o$ we have minimum extended$-s$ contribution 
(dotted line). Fig.3(d) illustrates how the extended$-s$ component evolves 
on the upper line of SC sites at either phase difference. 
Thus we conclude that the extended$-s$ 
and the $d-$wave components are in competition such that the 
extended$-s$ component is maximised at the detriment of the $d-$wave 
and vice versa. 

Fig.{\ref{fig4}} shows the calculated currents in both the 
local $s-$wave and non-local $d-$wave cases. To calculate the current (using 
Eq.5) we have to consider the flow across all possible routes in just 
one cell of our sample. In calculating currents it is essential to check 
current conservation: this is only guaranteed from a self-consistent 
solution. Here, we find conservation 
obeyed to within $0.01\%$. Consider the $s-$wave intially: the variation 
of current with phase-difference,$\varphi$, is 
plotted in Fig.4 (solid line)  where the values 
for phase difference range from $0^o$ to $+180^o$. 
It is immediately clear that the current is not sinusoidal in $\varphi$ 
but instead shows a sharp step at $180^o$. The remainder is roughly that 
of a saw-tooth albeit with some saturation. The step at $\pm 180^o$ can be 
attributed to resonant states entering the gap {\cite{Lick,Fuzaki}}. 
Fig.2(d) compares the local quasi-particle density of states at 
$\varphi=0^o and 180^o$ confirming the presence of resonant midgap 
states at $180^o$.

Our calculations for the $d$-wave current are presented in Fig.4 (dashed line). 
Again it is approximately a sawtooth. The slope $\frac{\partial I}{\partial \varphi}$ 
is also positive, and consequently this GB cannot be classified as a $\pi$-junction: 
this is consistent with the Sigrist-Rice formula for this 
geometry. Fogelstr{\"o}m and Yip{\cite{Fog} note that in certain geometries 
it is also possible to have a vanishing current at phase-differences other 
than integer multiples of $\pi$, and this they attribute to time reversal 
symmetry breaking (this has been reported in Il'ichev{\cite{Ill}). 
Fig.4 shows no such evidence and therefore we 
conclude that the symmetric grain-boundary does not have TRSB.

In conclusion we have developed a real-space method for determining how the 
order parameter and supercurrents change with phase-difference across a 
realistic interface in a superconductor. In this letter we have 
considered a large-angle symmetric tilt grain-boundary and considered the local 
$s-$wave and non-local $d-$wave pairing symmetry in the order-parameter on 
an equal footing. We have calculated the LDOS, Hartree-Fock terms, order 
parameter and current all self-consistently. By imposing a phase differnce, 
$\varphi$, across the junction we calculated the supercurrent $I(\varphi)$. 
We found, for both $s-$wave and $d-$wave that $I(\varphi)$ is non-sinusoidal but 
exhibits a saw-tooth like behaviour which can be attributed to a sudden 
filling-in of the energy gap at $\varphi=180^o$. Further, we note 
no time reversal symmetry breaking or $\pi$-junction behaviour 
in the $d-$wave case. 

We would like to thank B.L. Gyorffy for useful discussions throughout the 
course of this work. This work was supported by the EPSRC under grant number 
GR/L22454. 

%
%

\begin{figure} 
\caption{\it The symmetric model tilt grain-boundary. By periodicity, 
we carry out self-consistent calculations on two lines of sites (highlighted) 
which are then mirrored onto similar sites in the 
rest of the sample. Also shown is our definition of 
a $d$-wave $0^o$ phase difference across the boundary (see text).}
\label{fig1}
\end{figure}

\begin{figure}
\caption{\it a) order-parameter on lower line of SC sites at $0^o$ 
(full line) and $180^o$ (dotted line). The latter goes to zero 
on the percolation site ($x=0$),  
b)same as a) but for upper line of SC sites, c) self consistently determined 
evolution of the order-parameter phase 
on going through the GB for a $30^0$ phase difference, 
d) Quasiparticle local density of states at 
$0^o$ phase difference (full line) and at $180^o$ (dashed line).} 
\label{fig2}
\end{figure}

\begin{figure}
\caption{\it The contributions to the superconducting order-parameter 
($d-$wave) on going through the boundary. The x coordinate is plotted on the 
horizontal axis. Full line refers to $0^0$ phase difference, dashed refers 
to $180^0$: a) d-wave on lower line b) d-wave on upper line 
c) extended-s on lower line d) extended-s on upper line}
\label{fig3}
\end{figure}

\begin{figure}
\caption{\it The supercurent versus phase-difference in the order-parameter 
between the two bulk regions for the s-wave and d-wave cases.}
\label{fig4}
\end{figure}

%
%

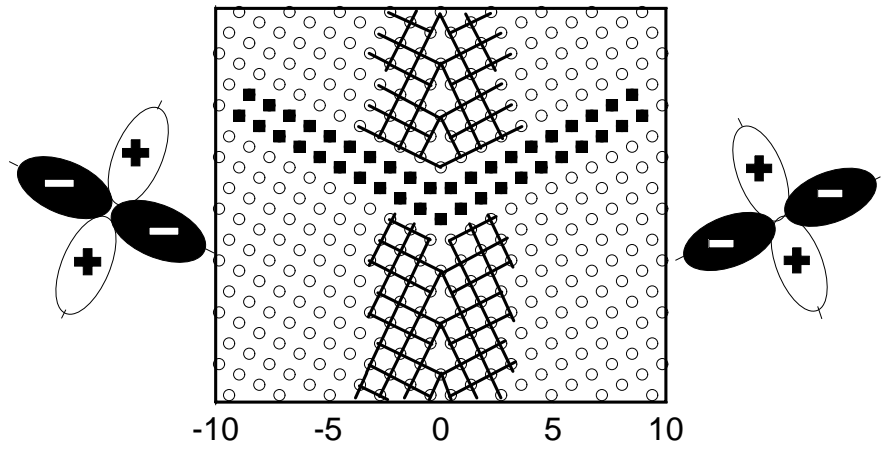
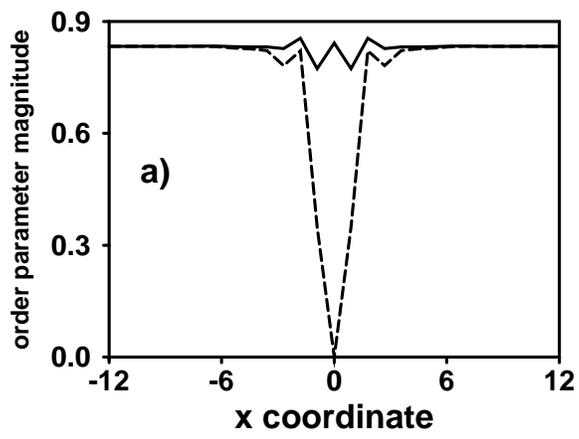
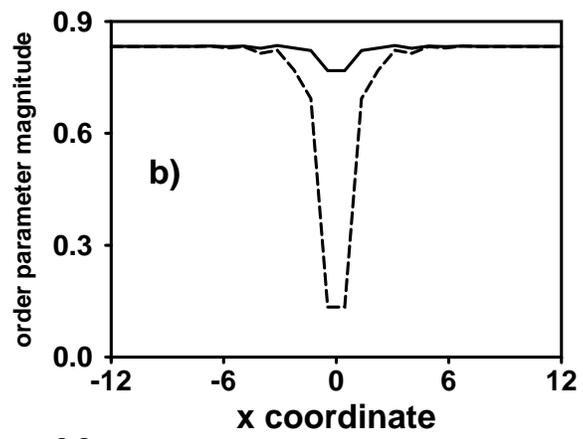
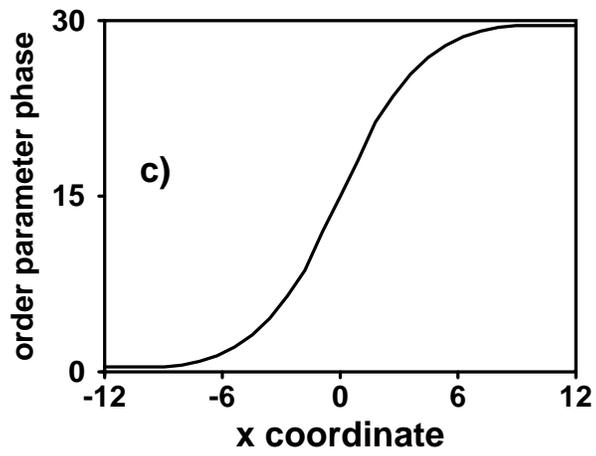
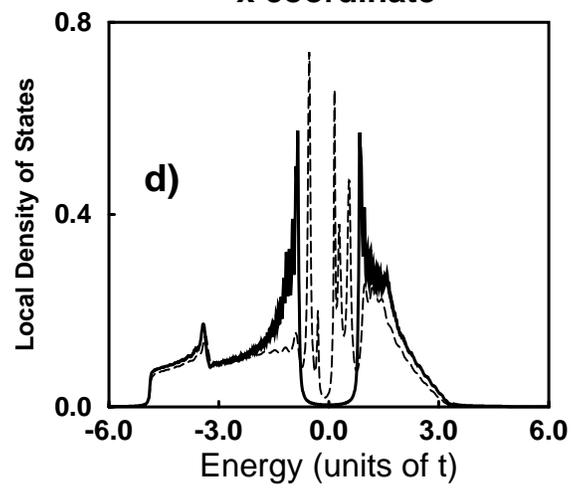
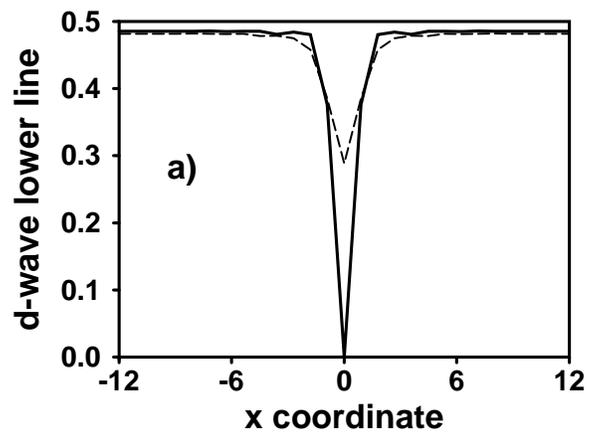
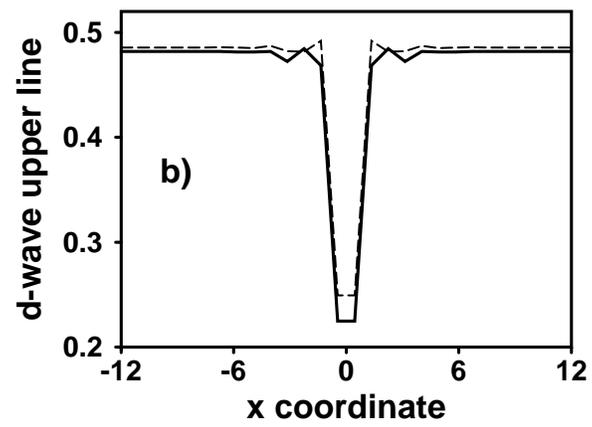
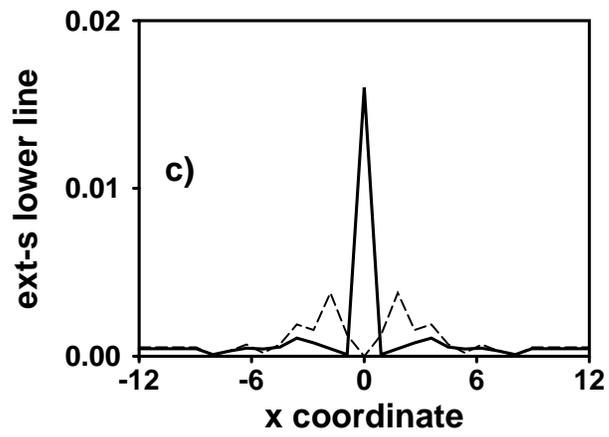
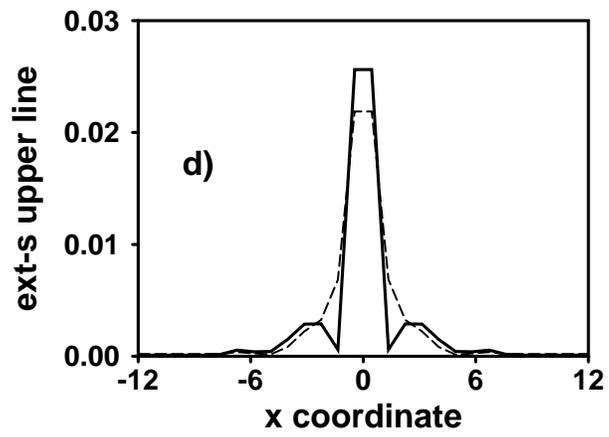
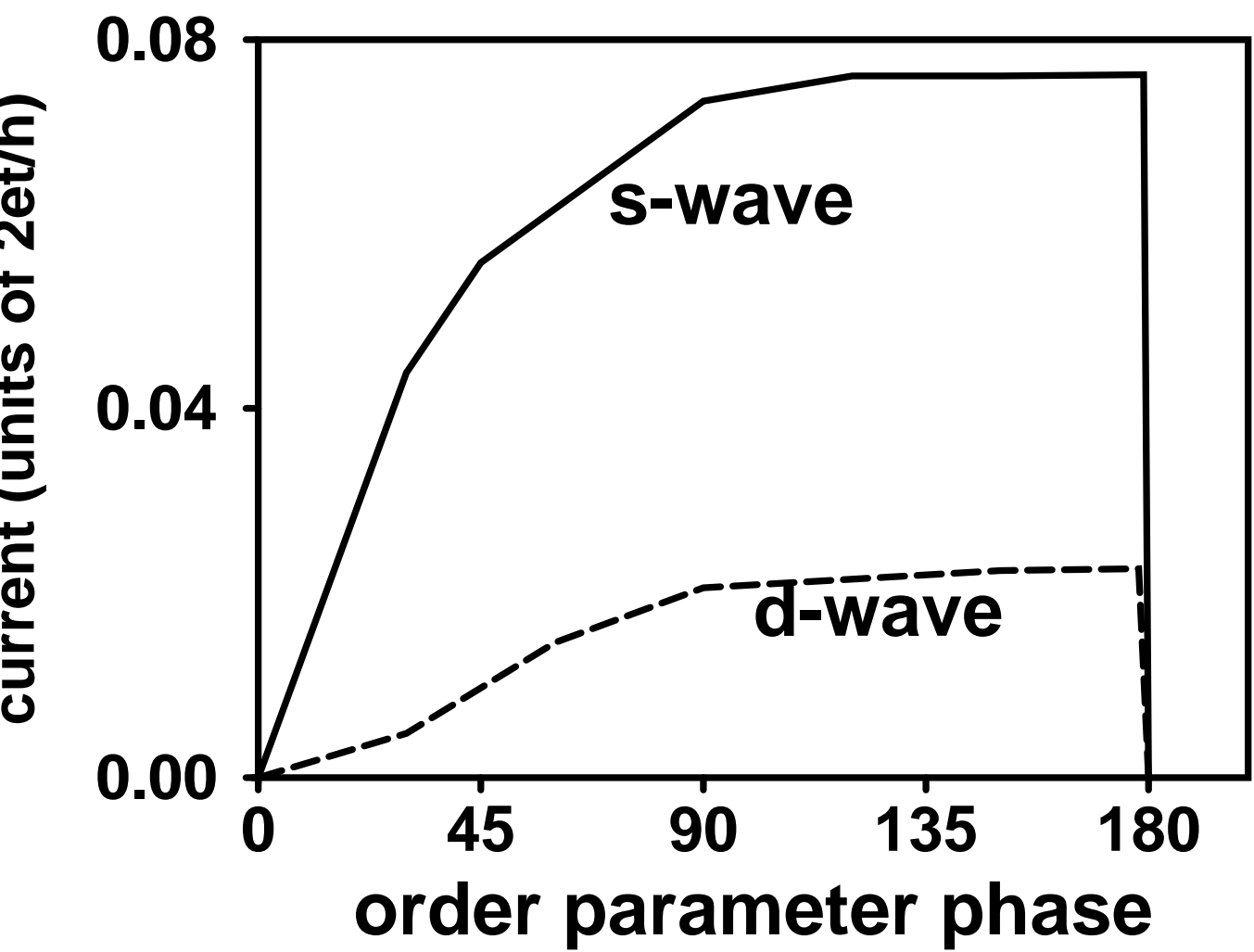
\end{document}